\documentclass[preprint,amsmath,amssymb,prd,superscriptaddress]{revtex4}
\usepackage[dvips]{graphicx}
\usepackage{color}
\usepackage{multirow}
\usepackage{array}
\usepackage{amsmath,bm}
\newcommand{\p}{\partial}
\newcommand{\pslash}{p\kern-1ex /}
\newcommand{\lslash}{l\kern-1ex /}
\newcommand{\kslash}{k\kern-1ex /}
\newcommand{\dslash}{\p\kern-1.2ex /}
\newcommand{\Dslash}{{\cal D}\kern-1.5ex /}
\newcommand{\Tr}{{\rm Tr}}

\newcommand{\re}{{\rm Re}}

\def\Id{\mbox{1\hspace{-1.2mm}I} }

\newcommand{\Dodwf}{\mathcal{D}}
\newcommand{\bea}{\begin{eqnarray}}
\newcommand{\eea}{\end{eqnarray}}

\newcommand{\nn}{\nonumber\\}

\newcommand{\BAN}{\begin{eqnarray*}}
\newcommand{\EAN}{\end{eqnarray*}}
\def\u{{\bf u}}
\def\d{{\bf d}}

\def\Id{ \mbox{1\hspace{-1.2mm}I} }
\begin{document}

\newcommand{\NTNU}{
  Physics Department, National Taiwan Normal University, Taipei, Taiwan~11677, R.O.C.
}

\newcommand{\ASIOP}{
  Institute of Physics, Academia Sinica, Taipei, Taiwan~11529, R.O.C. 
}

\newcommand{\NCTS}{
  Physics Division, National Center for Theoretical Sciences,
  National Tsing-Hua University, Hsinchu, Taiwan~30013, R.O.C.
}

\newcommand{\NTU}{
  Physics Department, National Taiwan University, Taipei, Taiwan~10617, R.O.C.
}

\newcommand{\CQSE}{
  Center for Quantum Science and Engineering,
  National Taiwan University, Taipei, Taiwan~10617, R.O.C. 
}

\preprint{NTUTH-19-505A}

\title{New $N_f=2$ Pseudofermion Action for Monte-Carlo Simulation of 
       Lattice Field Theory with Domain-Wall Fermions}

{

\author{Yu-Chih~Chen}
\noaffiliation

\author{Ting-Wai~Chiu}
\affiliation{\NTNU}
\affiliation{\ASIOP}
\affiliation{\NTU}

\noaffiliation


\begin{abstract}

We construct a novel $ N_f = 2 $ pseudofermion action for Monte-Carlo simulation of 
lattice gauge theory with domain-wall fermions (DWF), of which the  
effective four-dimensional lattice Dirac operator is equal to the 
overlap-Dirac operator with the argument of the sign function 
equal to $ H = c \gamma_5 D_w (1 + d D_w)^{-1} $, 
where $ c $ and $ d $ are parameters, and    
$D_w$ is the standard Wilson-Dirac operator plus a negative parameter $-m_0 \; (0 < m_0 < 2)$.
This new action is particularly useful for the challenging simulations 
of lattice gauge theories with large $N_f = 2n $ DWF, on the large lattices, 
and in the strong-coupling regime.   

\end{abstract}
\maketitle

\section{Introduction}

In lattice gauge theory with $N_f$ dynamical domain-wall fermions (DWF) \cite{Kaplan:1992bt},  
one often considers the cases containing two fermions with degenerate masses. 
For example, $N_f = 2 $, $N_f = 2+1$, and $N_f = 2+1+1 $
QCD, in which the masses of $\u$ and $\d$ quarks are degenerate, i.e., the theory 
in the isospin symmetry limit. For hybrid Monte-Carlo (HMC) simulations \cite{Duane:1987de}
of these theories, we have been using the $ N_f = 2 $ pseudofermion action 
as given in Ref. \cite{Chiu:2011bm}. 
Recently, for studies relating to the beyond Standard Model with composite Higgs boson, 
one investigates whether a lattice gauge theory with a large number of massless fermions  
can possess an infrared conformal fixed-point,  
e.g., the $SU(3)$ lattice gauge theory with $N_f=10$ massless fermions,   
in which the pseudofermion action can be written as 
the sum of five identical $N_f = 2 $ massless pseudofermion actions. 
However, the HMC simulation based on the traditional $N_f=2$ pseudofermion action 
turns out to be rather time-consuming for large lattices at strong couplings.
In this paper, we construct a new $ N_f=2 $ pseudofermion action for 
all variants of DWF, which is more efficient 
than the traditional $ N_f=2 $ pseudofermion action for the HMC simulation 
of the theories with large $N_f=2n$ massless DWF, especially for large lattices 
in the strong-coupling regime, as first reported in Ref. \cite{Chiu:2018edw}. 
In this paper, we demonstrate that, for the $SU(3)$ lattice gauge theory 
with $ N_f = 10 $ massless M\"obius DWF on the $ 32^4 $ lattice at $ 6/g_0^2 = 5.70 $,  
the efficiency (speed $\times$ acceptance rate) 
of the HMC simulation with the new action   
is about 1.2 times of its counterpart with the traditional action.  

In general, the 5-dimensional lattice Dirac operator of any kind of DWF
with infinite extent in the fifth dimension ($N_s = \infty$) gives 
the effective 4-dimensional lattice Dirac operator equal to 
\bea
\label{eq:Dmq}
D(m_q) = m_q + \frac{1}{2r} (1-rm_q) \left[ 1+\gamma_5 \frac{H}{\sqrt{H^2}} \right], 
\eea
where $ m_q $ is the bare quark mass,  
\bea
\label{eq:H}
H &=& c \gamma_5 D_w ( 1 + d D_w)^{-1}, \\
\label{eq:r}
r &=& \frac{1}{2 m_0 ( 1 - d m_0)}, 
\eea
and $D_w$ is the standard Wilson Dirac operator plus a negative parameter $-m_0 \; (0 < m_0 < 2)$.
Here $ c $ and $ d $ are parameters to specify the type of DWF.   
Setting $ m_q = 0 $, $c=1$ and $ d=0 $, (\ref{eq:Dmq}) becomes   
the massless overlap-Dirac operator \cite{Neuberger:1997fp,Narayanan:1994gw}, 
\BAN
\label{eq:Dov}
D_{ov} = m_0 \left[ 1+ D_w (D_w^\dagger D_w)^{-1/2} \right] 
       = m_0 \left[ 1+ \gamma_5 \frac{H_w}{\sqrt{H_w^2}} \right], \hspace{4mm} H_w = \gamma_5 D_w.   
\EAN
In other words, the effective 4-dimensional lattice Dirac operator (\ref{eq:Dmq}) of any kind of DWF
can be regarded as the generalized overlap-Dirac operator with the argument of 
the sign function equal to $ H $ (\ref{eq:H}). In the massless limit $m_q =0 $, 
(\ref{eq:Dmq}) satisfies the Ginsparg-Wilson relation \cite{Ginsparg:1981bj}
\BAN
D \gamma_5 + \gamma_5 D = 2 r D \gamma_5 D \Longleftrightarrow  
D^{-1} \gamma_5  +  \gamma_5 D^{-1} = 2 r \gamma_5 \Id,     
\EAN
where the chiral symmetry is broken by a contact term, 
i.e., the exact chiral symmetry at finite lattice spacing. 

On the 5-dimensional lattice with size ($N_x^3 \times N_t \times N_s $), 
the lattice fermion operator 
of all variants of DWF \cite{Shamir:1993zy, Borici:1999zw, Chiu:2002ir, Brower:2004xi} can be written as   
\bea
\label{eq:D_odwf}
[\Dodwf(m_q)]_{xx';ss'} &=&
  (\rho_s D_w + 1)_{xx'} \delta_{ss'}
 +(\sigma_s D_w - 1)_{xx'} L_{ss'},
\eea
where $ s $ and $ s' $ are the indices in the fifth dimension, 
$ x $ and $ x' $ denote the lattice sites on the 4-dimensional lattice, 
and the Dirac and color indices have been suppressed.
Here $D_w$ is the standard Wilson Dirac operator plus a negative parameter $-m_0 \; (0 < m_0 < 2)$,
%
\BAN
(D_w)_{xx'} = (4-m_0) -\frac{1}{2} \sum_{\hat\mu=1}^4 \left[
  (1-\gamma_\mu)U_\mu(x)\delta_{x+\hat{\mu},x'}
 +(1+\gamma_\mu)U^\dagger_\mu(x')\delta_{x-\hat{\mu},x'} \right],
\EAN
where $U_\mu(x)$ denotes the link variable pointing from $ x $ to $ x + \hat\mu $.
The operator $ L $ is independent of the gauge field, and it can be written as
\bea
\label{eq:L}
L = P_+ L_+ + P_- L_-, \quad P_\pm = (1\pm \gamma_5)/2,
\eea
where
\bea
\label{eq:Lpm}
(L_+)_{ss'} = \left\{
    \begin{array}{ll}
      - r m_q \delta_{N_s,s'}, & s = 1,  \\
      \delta_{s-1,s'}, & 1 < s \leq N_s
    \end{array}\right.,
\quad L_-=(L_+)^{T}.
\eea
Note that the matrices $ L_{\pm} $ satisfy $ L_\pm^T = L_\mp $, and $ R_5 L_\pm R_5 = L_\mp $,
where $ R_5 $ is the reflection operator in the fifth dimension, 
with elements $ (R_5)_{ss'} = \delta_{s',N_s+1-s} $.
Thus $ R_5 L_\pm $ is real and symmetric. 

For all variants of DWF in Refs. \cite{Shamir:1993zy, Borici:1999zw, Chiu:2002ir, Brower:2004xi},    
we write $ \rho_s = c \omega_s + d $, and $ \sigma_s = c \omega_s - d $,   
where $ c $, $ d $, and $\{ \omega_s \} $  are parameters to specify the type of DWF.
For the conventional DWF with the Shamir kernel \cite{Shamir:1993zy}, 
$c=d=1/2 $, and $ \omega_s = 1, \forall s $.
For the Borici DWF \cite{Borici:1999zw}, $ c=1 $, $ d=0 $, and $ \omega_s = 1, \forall s $.
For the optimal DWF \cite{Chiu:2002ir}, the weights $ \{ \omega_s \} $ are fixed according 
to the formula derived in \cite{Chiu:2002ir} such that the maximal chiral symmetry is attained, 
i.e., its effective 4-dimensional Dirac operator is exactly equal to the 
Zolotarev optimal rational approximation 
of (\ref{eq:Dmq}).
For the M\"obius DWF \cite{Brower:2004xi}, $ \omega_s = 1, \forall s $.

Using the lattice DWF operator (\ref{eq:D_odwf}), and including the Pauli-Villars fields with bare mass
$ m_{PV} = 1/r = 2 m_0 ( 1 - d m_0) $, 
the pseudofermion action for all variants of DWF can be written as
\bea
\label{eq:S_nf1}
S = \phi^\dagger \frac{\Dodwf(m_{PV})}{\Dodwf(m_q)} \phi, 
\eea
where $ \phi $ and $ \phi^\dagger $ are complex scalar fields carrying the same quantum numbers
(color, spin) of the fermion fields. 
Integrating the pseudofermion fields in the fermionic partition function 
gives the fermion determinant of the effective 4-dimensional lattice Dirac operator at finite $ N_s $
\BAN
\label{eq:Z_nf1}
\int [d\phi^{\dagger}][d\phi] \exp\left\{ -\phi^\dagger \frac{\Dodwf(m_{PV})}{\Dodwf(m_q)} \phi \right\}
= \det \frac{\Dodwf(m_q)}{\Dodwf(m_{PV})} 
= \det  D_{N_s}(m_q),    
\EAN
where 
\BAN
\label{eq:odwf_4d}
\begin{aligned}
D_{N_s}(m_q) &= m_q + \frac{1}{2r}(1-rm_q) [1+ \gamma_5 S_{N_s}(H) ], \\
S_{N_s}(H) &= \frac{1-\prod_{s=1}^{N_s} T_s}{1 + \prod_{s=1}^{N_s} T_s}, \hspace{4mm}  
T_s = \frac{1-\omega_s H}{1+ \omega_s H}.
\end{aligned}
\EAN
In the limit $ N_s \to \infty $, $ S_{N_s}(H) \to H/\sqrt{H^2} $, thus $ D_{N_s}(m_q) \to D(m_q) $ 
in (\ref{eq:Dmq}). For finite $N_s$, if the weights $ \{ \omega_s, s = 1, \cdots, N_s \} $ 
are fixed according to the formulas derived in Ref. \cite{Chiu:2002ir},   
$ S_{N_s}(H) $ is equal to the Zolotarev optimal rational approximation of $ H/\sqrt{H^2} $.
On the other hand, if the weights $ \omega_s = 1, \forall s $, 
$ S_{N_s}(H) $ is equal to the polar approximation of $ H/\sqrt{H^2} $.

Note that the pseudofermion action (\ref{eq:S_nf1}) for $ N_f = 1 $ DWF cannot be used for HMC simulations, 
since $ \Dodwf(m_{PV}) \Dodwf^{-1}(m_q) $ is not positive-definite and Hermitian.  
A positive-definite, Hermitian, and exact pseudofermion action for $N_f =1 $ DWF 
has been constructed in Ref. \cite{Chen:2014hyy}. For $ N_f=2$, it is straightforward 
to construct a positive-definite and Hermition pseudofermion action from (\ref{eq:S_nf1}), 
\BAN
\label{eq:S_nf2}
S^{N_f=2} = \phi^\dagger \Dodwf^\dagger(m_{PV}) [\Dodwf(m_q) \Dodwf^\dagger(m_q)]^{-1} \Dodwf(m_{PV}) \phi.  
\EAN
However, this $ N_f = 2 $ pseudofermion action is not efficient for the HMC simulation.

The outline of this paper is as follows. In section 2, we derive the  
traditional $ N_f = 2 $ pseudofermion action for DWF, which we have been using for the simulations of 
$ N_f=2 $ and $ N_f = 2+1+1 $ lattice QCD. Even though the derivation of this $ N_f = 2 $ 
pseudofermion action has been given in Ref. \cite{Chiu:2013aaa}, we present  
a different derivation here, mainly for defining our notations in this paper. 
In section 3, we construct the new $ N_f = 2 $ pseudofermion action for all variants of DWF. 
In section 4, we perform numerical simulations to compare the HMC efficiencies 
of the new and the old actions. 
In section 5, we conclude with some remarks.

\section{The traditional $N_f=2$ pseudofermion action}
 
Using $ \rho_s = c \omega_s + d $, and $ \sigma_s = c \omega_s - d $, (\ref{eq:D_odwf}) can be rewritten as 
\BAN
\Dodwf(m_q) &=& D_w [ c \omega (1+L) + d (1 -L)] + 1 - L,  
\EAN
where $ L $ is defined in (\ref{eq:L}) and (\ref{eq:Lpm}).
Since $ \det [\Dodwf(m_q)] = \det [ \omega^{-1/2} \Dodwf(m_q) \omega^{1/2} ] $, 
we can use $ \omega^{-1/2} \Dodwf(m_q) \omega^{1/2} $ for the HMC simulation.  
Moreover, $ L $ and $\omega = \mbox{diag}(\omega_1, \cdots, \omega_{N_s})$ 
are independent of the gauge field, we can drop the factor 
$ [c \omega^{1/2} (1+L) \omega^{1/2} + d \omega^{-1/2} (1-L) \omega^{1/2} ] $ 
from $ \omega^{-1/2} \Dodwf(m_q) \omega^{1/2} $, and obtain the re-scaled DWF operator for HMC,
\bea
\label{eq:DT}
D_{T}(m_q)
= D_w + \omega^{-1/2}\left\{c[1+L(m_q)][1-L(m_q)]^{-1} + d \omega^{-1} \right\}^{-1}\omega^{-1/2}  
= D_w + M(m_q),
\eea
where  
\bea
\label{eq:M}
M(m_q) &=& P_+ M_+(m_q) + P_- M_-(m_q), \quad P_\pm = (1\pm \gamma_5)/2, \\
\label{eq:M_pm}
M_\pm(m_q) &=& \omega^{-1/2} \left\{c N_{\pm}(m_q) + d \omega^{-1} \right\}^{-1} \omega^{-1/2}, \\
\label{eq:N_pm}
N_{\pm}(m_q) &=& [1+L_{\pm}(m_q)][1-L_{\pm}(m_q)]^{-1}.
\eea
Here the dependence on $ m_q $ has been shown explicitly in $ L_{\pm} $, $ M_\pm $, and $ N_\pm $.

Next, we perform the even-odd preconditioning on $D_T(m_q)$.
This is essential for halving the memory footprint as well as 
lowering the condition number of the conjugate gradient solver for the fermion force. 
With even-odd preconditioning on the 4-dimensional space-time lattice, 
(\ref{eq:DT}) can be written as 
\bea
D_T(m_q) =
\begin{pmatrix}
M_5^{-1}(m_q) & D_w^{\text{EO}} \\
D_w^{\text{OE}} &  M_5^{-1}(m_q)
\end{pmatrix} 
\label{eq:DT_eo} 
\eea
where
\begin{equation}
M_5^{-1}(m_q) = 4 - m_0 + M(m_q). 
\end{equation}
Using the Schur decomposition, (\ref{eq:DT_eo}) becomes
\begin{equation}
D_T(m_q) =
\begin{pmatrix}
1 & 0 \\
D_w^{\text{OE}} M_5 & 1
\end{pmatrix}
\begin{pmatrix}
M_5^{-1} & 0 \\
0 & M_5^{-1} C
\end{pmatrix}
\begin{pmatrix}
1 & M_5 D_w^{\text{EO}} \\
0 & 1
\end{pmatrix}
\label{eq:D_odwf_decomp}
\end{equation}
where 
\begin{equation}
\label{eq:C_def}
C(m_q) \equiv 1 - M_5(m_q) D_w^{\text{OE}} M_5(m_q) D_w^{\text{EO}}.
\end{equation}
Since $ \det D_T = \det M_5^{-2} \cdot \det C $, and
$ M_5 $ does not depend on the gauge field, we can just use $ C $
in the Monte Carlo simulation. After including the Pauli-Villars fields with mass 
$ m_{PV} = 2m_0(1-d m_0)$, we obtain the $ N_f=2 $ pseudofermion action for all variants of DWF,  
\bea
\label{eq:S_old2F}
S = \phi^\dagger C^\dagger(m_{PV}) \{ C(m_q) C^\dagger(m_q) \}^{-1} C(m_{PV}) \phi, 
\eea
which has been used for the HMC simulations of $ N_f = 2 $, 
and $ N_f = 2+1+1 $ lattice QCD with the optimal DWF \cite{Chen:2014hva,Chen:2017kxr,Chiu:2018qcp}.

\section{The new $N_f=2$ pseudofermion action}


From (\ref{eq:DT}), it gives the ratio 
\bea
\label{eq:DT_ratio}
\frac{\det[D_T(m_{PV})]}{\det[D_T(m_q)]} &=& \frac{\det[D_w+M(m_{PV})]}{\det[ D_w+M(m_q)]} \nn
&=&\frac{\det[ H_T(m_q)+\gamma_5 \Delta(m_q)]}{\det[ H_T(m_q)]}
    =\det\left[1 +\gamma_5 \Delta(m_q)\frac{1}{H_T(m_q)}\right], 
\eea
where $ \det R_5 $ and $ \det \gamma_5 $ have been multiplied in both the numerator and the denominator, and 
\BAN
H_T(m_q) &=& R_5 \gamma_5 D_T(m_q), \\
\Delta(m_q) &=& R_5 \left[ M(m_{PV})-M(m_q) \right] = P_+ \Delta_+(m_q) + P_- \Delta_-(m_q).
\EAN
Following Ref. \cite{Chen:2014hyy}, $ \Delta_\pm(m_q) $ can be simplified as follows.
From (\ref{eq:N_pm}), it gives the relation
\BAN
N_\pm(m_q) = N_\pm(0) - \frac{2 r m_q}{1+ r m_q} u u^T, \quad u^{T} \equiv (1,1,\cdots,1).
\EAN
Then using the Sherman-Morrison formula, we obtain
\bea
\label{eq:Mpm}
[cN_{\pm}(m_q) + d \omega^{-1} ]^{-1}
= A_{\pm}^{-1} + \frac{2c r m_q}{1+ r m_q - 2 c r m_q \lambda_\pm } A_{\pm}^{-1}uu^{T} A_{\pm}^{-1},
\eea
where
\BAN
A_{\pm} &=& cN_{\pm}(0) + d \omega^{-1} , \\
\lambda_\pm &=& u^T A_\pm^{-1} u.
\EAN
Now using the optimal $\omega$ which is invariant 
under $ R_5 $ (i.e., $ R_5 \omega R_5 = \omega $) \cite{Chiu:2015sea}, 
defining $v_{\pm} \equiv R_5 A_{\pm}^{-1} u$, and putting (\ref{eq:Mpm}) into (\ref{eq:M_pm}), 
we obtain 
\bea
\label{eq:Mpm_vvT}
M_{\pm}(m_q) = \omega^{-1/2}A_{\pm}^{-1}\omega^{-1/2} +
      \frac{2c r m_q}{1+r m_q-2cr m_q \lambda_\pm } R_5 \omega^{-1/2} v_{\pm}v_{\pm}^{T}\omega^{-1/2}, 
\eea
where we have used  
\BAN 
u^{T} A_{\pm}^{-1}= u^{T}R_5 R_5 A_{\pm}^{-1}R_5 R_5 = u^{T} (A_{\pm}^{-1})^{T} R_5 
                  = (R_5 A_{\pm}^{-1}u)^{T} = v_\pm^T.  
\EAN
Since $ A_{\pm}^{-1} $ is an lower/upper triangular matrix, we can solve for $ v_{\pm} $ 
(vectors in the fifth dimensional space) with the following recursion relation,
\BAN
(v_+)_{N_s} &=& (v_-)_{1} = \alpha_{N_s},\\
(v_+)_s &=& (v_-)_{N_s-s+1} = \alpha_s \beta_{s+1} (v_+)_{s+1}, \quad s=N_s-1, \cdots, 1,
\EAN
where $\alpha_s = 1/(c+ d \omega_s^{-1})$ and $\beta_s = -c + d \omega_s^{-1}$.
Then we obtain
\BAN
\lambda_{-} = \lambda_{+} = u^{T}A_{+}^{-1}u = u^{T} R_5 A_{+}^{-1}u = u^{T}v_{+} = \sum_s (v_+)_s = \sum_s \alpha_s Q_s \equiv \lambda,
\EAN
where $ Q_s \equiv \alpha_{s+1}\beta_{s+1}...\alpha_{Ns}\beta_{Ns} $.
Using (\ref{eq:Mpm_vvT}), we obtain
\bea
\label{eq:Delta_pm}
\Delta_\pm(m_q) = R_5 \left[ M_\pm(m_{PV})-M_\pm(m_q) \right] 
                = k \omega^{-1/2} v_{\pm}v_{\pm}^{T} \omega^{-1/2},  
\eea
where
\BAN
k = \frac{c}{1-c\lambda}\left(\frac{1-r m_q}{1+r m_q-2c r m_q \lambda}\right), \hspace{4mm} 
r = \frac{1}{2 m_0 ( 1 - d m_0)}. 
\EAN
Putting (\ref{eq:Delta_pm}) into (\ref{eq:DT_ratio}) and using the identity $\det(1+AB) = \det(1+BA)$, 
(\ref{eq:DT_ratio}) becomes  
\BAN
\frac{\det[D_T(m_{PV})]}{\det[D_T(m_q)]}
=\det\left[ 1 + k \gamma_5 v^{T}\omega^{-1/2}\frac{1}{H_T(m_q)}\omega^{-1/2}v \right]
=\det K(m_q)
\EAN
where
\bea
\label{eq:K}
K(m_q) = 1 + k \gamma_5 v^{T}\omega^{-1/2}\frac{1}{H_T(m_q)}\omega^{-1/2}v, \hspace{4mm}
v = P_+ v_+ + P_- v_-.
\eea
Note that $ K(m_q) $ is an operator on the 4-dimensional space, similar to the 
positive-definite Hermitian operators $ H_1 $ and $ H_2 $ in the exact pseudofermion action 
for one-flavor DWF \cite{Chen:2014hyy}.
However, $K(m_q)$ is not Hermitian. For the $N_f=2 $ pseudofermion action, it can be constructed as
\bea
\label{eq:S_new2F}
S_{\rm new} = \phi^{\dagger} K^{\dagger}(m_q) K(m_q) \phi, 
\eea
where $ \phi $ and $ \phi^\dagger $ are pseudofermion fields on the 4-dimensional lattice.
This is the main result of this paper.

\section{Numerical Results}

To compare the efficiencies and characteristics of the HMC simulations with the 
new action and the traditional action, we focus on the challenging simulations 
of lattice gauge theories with large $N_f = 2n $ massless DWF on a large lattice. 
Here we perform HMC simulations of the $SU(3)$ lattice gauge theory with $N_f=10$ massless fermions
on the $ 32^4 $ lattice, using the M\"obius DWF ($c=1$, $ d=1/2$ and $\omega_s = 1, \forall s $) 
with $ N_s = 16 $ and $ m_0 = 1.8 $, and the Wilson plaquette gauge action at $ 6/g_0^2 = 5.70 $. 
Since to simulate $ N_f = 10 $ fermions amounts to simulate 5 pairs of $ N_f = 2 $ fermions, 
for the traditional $N_f=2$ pseudofermion action (\ref{eq:S_old2F}), the partition function 
of the $ SU(3) $ gauge theory with $ N_f = 10 $ massless DWF can be written as
\bea
\label{eq:Z_old2F}
Z = \int[dU]\prod_{i=1}^5 [d\phi_i^{\dag}][d\phi_i]
\exp \left( -S_g[U]- \sum_{i=1}^5 \phi_i^\dagger 
            C_i^\dagger(m_{PV}) [ C_i(0) C_i^\dagger(0)]^{-1} C_i(m_{PV}) \phi_i \right),
\eea
where $ \phi_i $ and $ \phi_i^\dagger $ are pseudofermion fields, 
$m_{PV} = m_0(2-m_0) = 0.36 $ is the mass of the Pauli-Villars fields,   
and $ S_g(U) $ is the Wilson plaquette gauge action
\BAN
S_g(U) = \frac{6}{g_0^2} \sum_{plaq.}\left\{1-\frac{1}{3} \re \Tr (U_p) \right\}.
\EAN
Similarly, using the new $N_f=2$ pseudofermion action (\ref{eq:S_new2F}),  
the partition function of the $SU(3)$ lattice gauge theory with $N_f=10$ massless fermions 
can be written as
\bea
\label{eq:Z_new2F}
Z_{\text{new}} = \int[dU]\prod_{i=1}^5 [d\phi_i^{\dag}][d\phi_i]
\exp \left(-S_g[U]- \sum_{i=1}^5 \phi^\dagger_i K(0)^\dagger K(0) \phi_i \right).
\eea
In both cases, the boundary conditions of the gauge field are periodic in all directions,
while the boundary conditions of the pseudofermion fields are antiperiodic in all directions.
In the molecular dynamics, we use the Omelyan integrator \cite{Omelyan:2001},
and the Sexton-Weingarten multiple-time scale method \cite{Sexton:1992nu}.
Moreover, we introduce auxiliary heavy fermion fields (with mass $ m_H a = 0.1 $)  
for the mass-preconditioning, similar to the case of Wilson fermion \cite{Hasenbusch:2001ne}.
Each simulation is performed on a single Nvidia GTX-TITAN.
The thermalization for the HMC simulation with the traditional/new action 
takes 164/172 trajectories, then to generate 78/79 trajectories after thermalization. 
The total computation time for the case with the traditional action takes about 
58 days, while that with the new action about 50 days.
The HMC characteristics based on the thermalized trajectories 
are summarized in Table \ref{tab:HMC_summary}.

\begin{table}[h!]
\begin{center}
\caption{Summary of the HMC characteristics for the $SU(3)$ gauge theory with $N_f=10$ massless DWF. }
\setlength{\tabcolsep}{2pt}
\vspace{2mm}

\begin{tabular}{|c|cccccc|}
\hline
  & Time(s)/traj 
  & Acceptance  
  & $ \left< \Delta H \right> $ 
  & $P_{\rm acc} = {\rm erfc}(\sqrt{\left< \Delta H \right>}/2)$ 
  & $ \left< \exp(-\Delta H) \right> $ 
  & $ \left<{\rm plaquette} \right> $ \\
\hline
Traditional & 20455(40) & 0.846(41) & 0.053(34) & 0.871(44) & 1.011(35) & 0.57381(2)\\
\hline
New & 16992(36) & 0.823(43) & 0.110(47) & 0.814(40) & 0.983(54) & 0.57384(2) \\
\hline
\end{tabular}
\label{tab:HMC_summary}
\end{center}
\end{table}

\begin{figure*}[h!]
\begin{center}
\begin{tabular}{@{}c@{}c@{}}
\includegraphics*[width=8cm,clip=true]{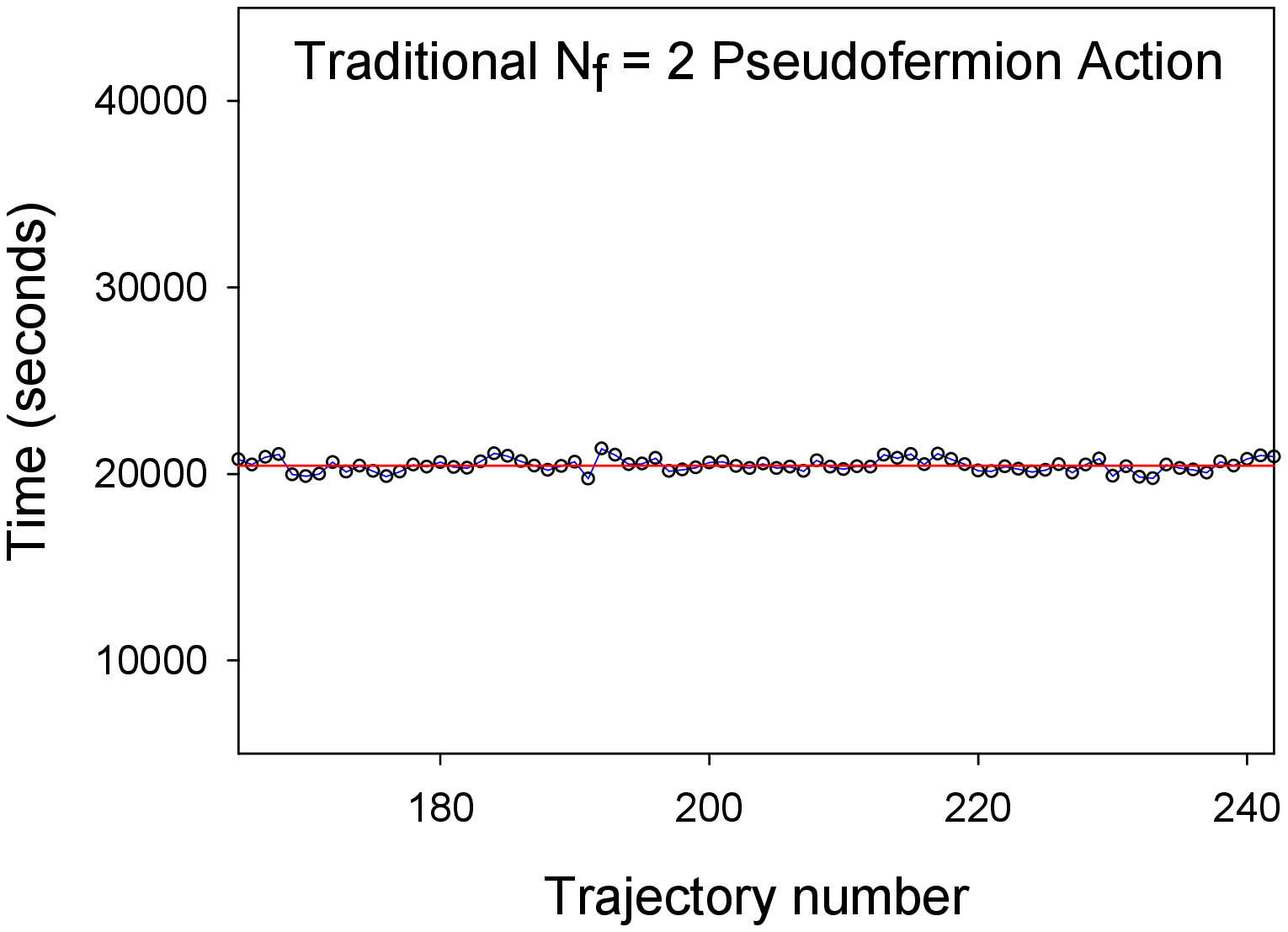}
&
\includegraphics*[width=8cm,clip=true]{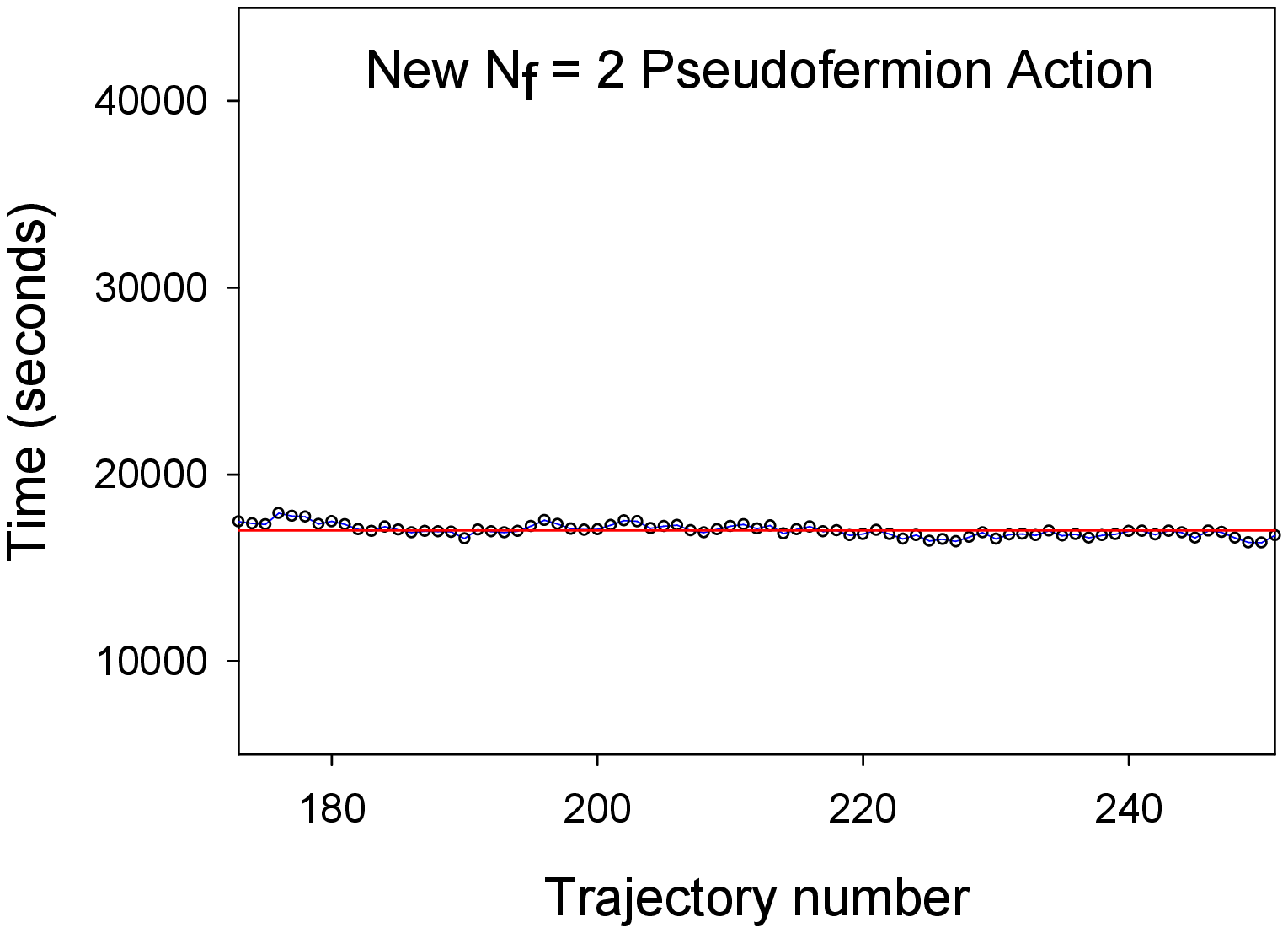}
\\ (a) & (b)
\end{tabular}
\caption{The computational time for each trajectory in the HMC simulation
of the $SU(3)$ lattice gauge thoery with $ N_f=10 $ massless M\"obius DWF, 
(a) the traditional action, and (b) the new action. The horizontal line 
is the average computational time per trajectory.   
The line connecting the data points is only for guiding the eyes.}
\label{fig:HMC_time}
\end{center}
\end{figure*}

In Fig. \ref{fig:HMC_time}, we plot the computational time for each HMC trajectory 
in the HMC simulations of the $SU(3)$ lattice gauge thoery with $ N_f=10 $ massless M\"obius DWF. 
From the first column of Table \ref{tab:HMC_summary}, the average speed of the 
simulation with the new action is more than 1.2 times of its counterpart with the traditional action. 
Moreover, from the second column of Table \ref{tab:HMC_summary},
the acceptance rate with the new action is 0.823(43), compatible with   
its counterpart with the traditional action, 0.846(41). 
Taking into account of both the speed and the acceptance rate, 
the HMC efficiency (speed $\times$ acceptance rate) 
with the new action is estimated to be about 1.2 times of 
its counterpart with the traditional action. 
In the following, we present more details of the simulations.

\begin{figure*}[h!]
\begin{center}
\begin{tabular}{@{}c@{}c@{}}
\includegraphics*[width=8cm,clip=true]{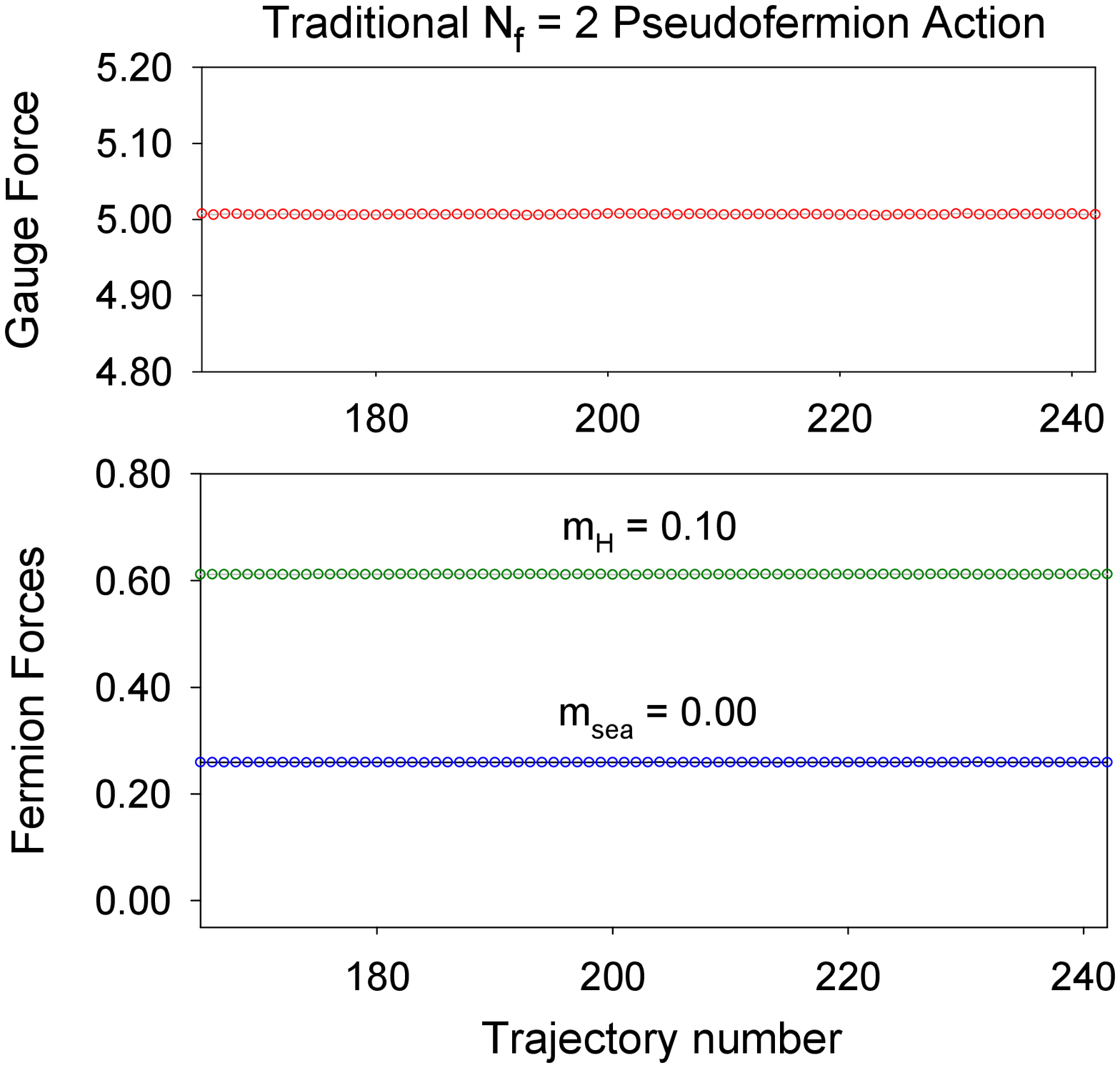}
&
\includegraphics*[width=8cm,clip=true]{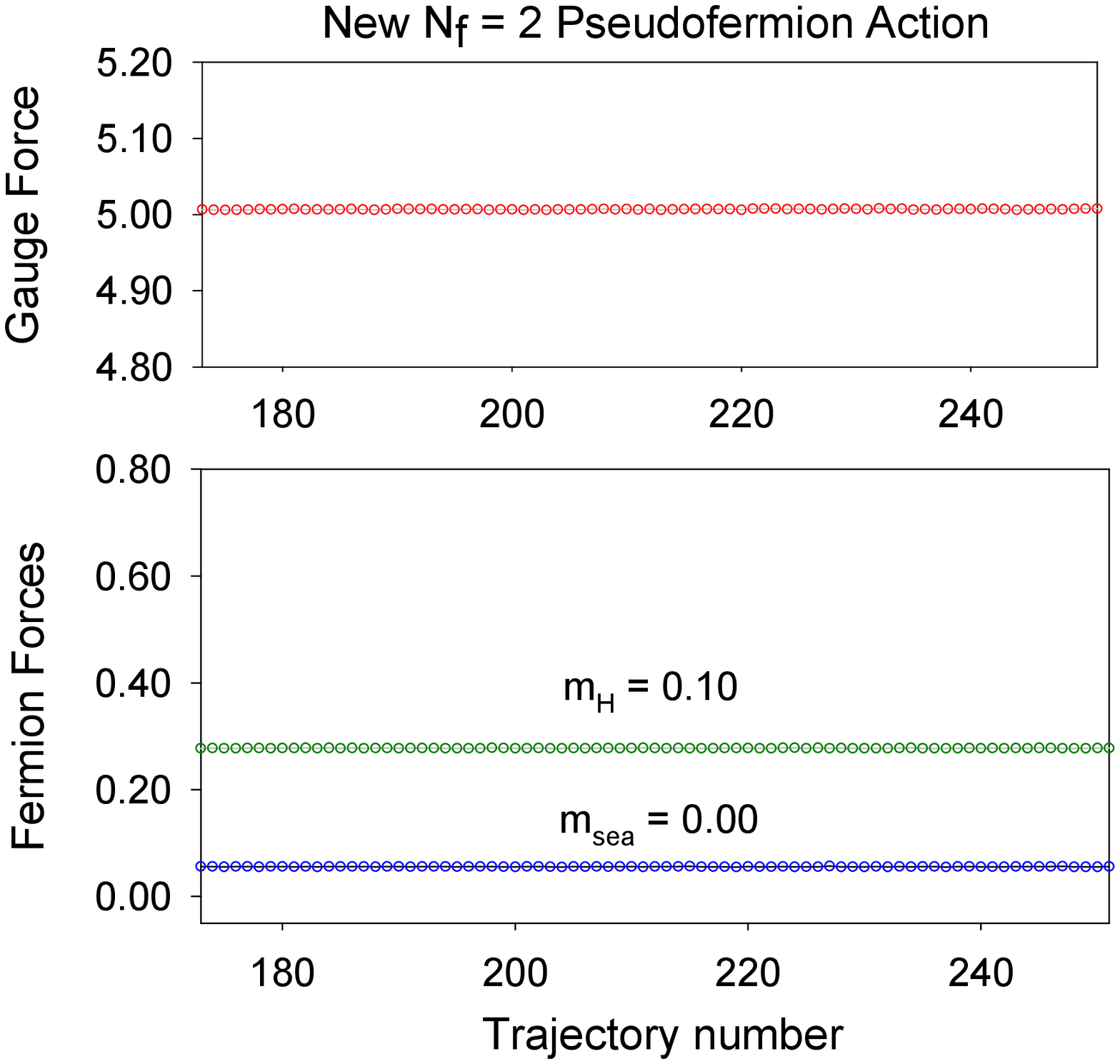}
\\ (a) & (b)
\end{tabular}
\caption{The maximum forces of the gauge field, the heavy fermion field, and the light fermion field
versus the HMC trajectory in the HMC simulations of  
the lattice gauge thoery with $ N_f=10 $ massless M\"obius DWF,   
(a) the traditional action, and (b) the new action.
Here only the fermion forces corresponding to the first pair of pseudofermions are plotted, i.e., 
for $ i=1 $ in (\ref{eq:Z_old2F}) and (\ref{eq:Z_new2F}).
}
\label{fig:HMC_forces}
\end{center}
\end{figure*}

In Fig. \ref{fig:HMC_forces}, we plot the maximum force (averaged over all links)
among all momentum updates in each trajectory, for the gauge field, the heavy fermion field,
and the light fermion field respectively. 
Here only the fermion forces corresponding to the first pair of pseudofermions are plotted, i.e., 
for $ i=1 $ in (\ref{eq:Z_old2F}) and (\ref{eq:Z_new2F}). The fermion forces for 
$ i=2,\cdots 5 $ behave similar to those of $ i=1 $, thus are omitted.     
For both (a) and (b), the forces all behave smoothly for all trajectories. 
However, the fermion forces in (b) with the new action are substantially 
smaller than their counterparts in (a) with the traditional action.
This immediately implies that the step sizes (in the Omelyan integrator) 
for the momentum update with the fermion forces in the case with the new action 
can be larger than their counterparts with the traditional action. 
With the length of the HMC trajectory equal to one, we use 3 different time scales 
for the momentum update with the gauge force, the heavy-fermion force, and the light-fermion force,  
which correspond to the step sizes $1/(k_0 k_1 k_2)$, $1/(k_1 k_2)$, and $1/k_2$ respectively. 
In our simulations, we set $(k_0, k_1, k_2)=(10, 2, 14)$ for the traditional action, 
while $(k_0, k_1, k_2)=(10, 2, 5)$ for the new action.

\begin{figure*}[h!]
\begin{center}
\begin{tabular}{@{}c@{}c@{}}
\includegraphics*[width=8cm,clip=true]{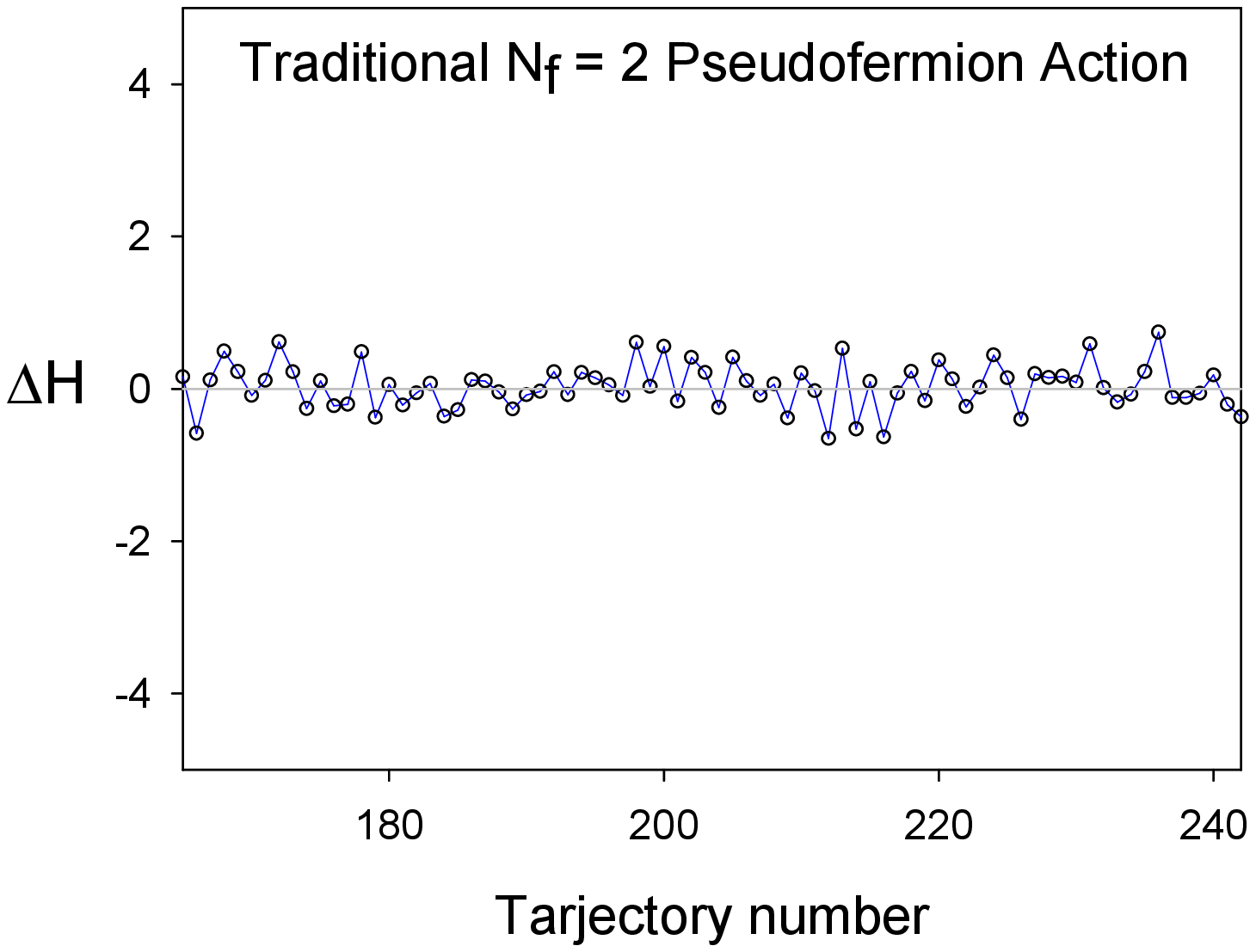}
&
\includegraphics*[width=8cm,clip=true]{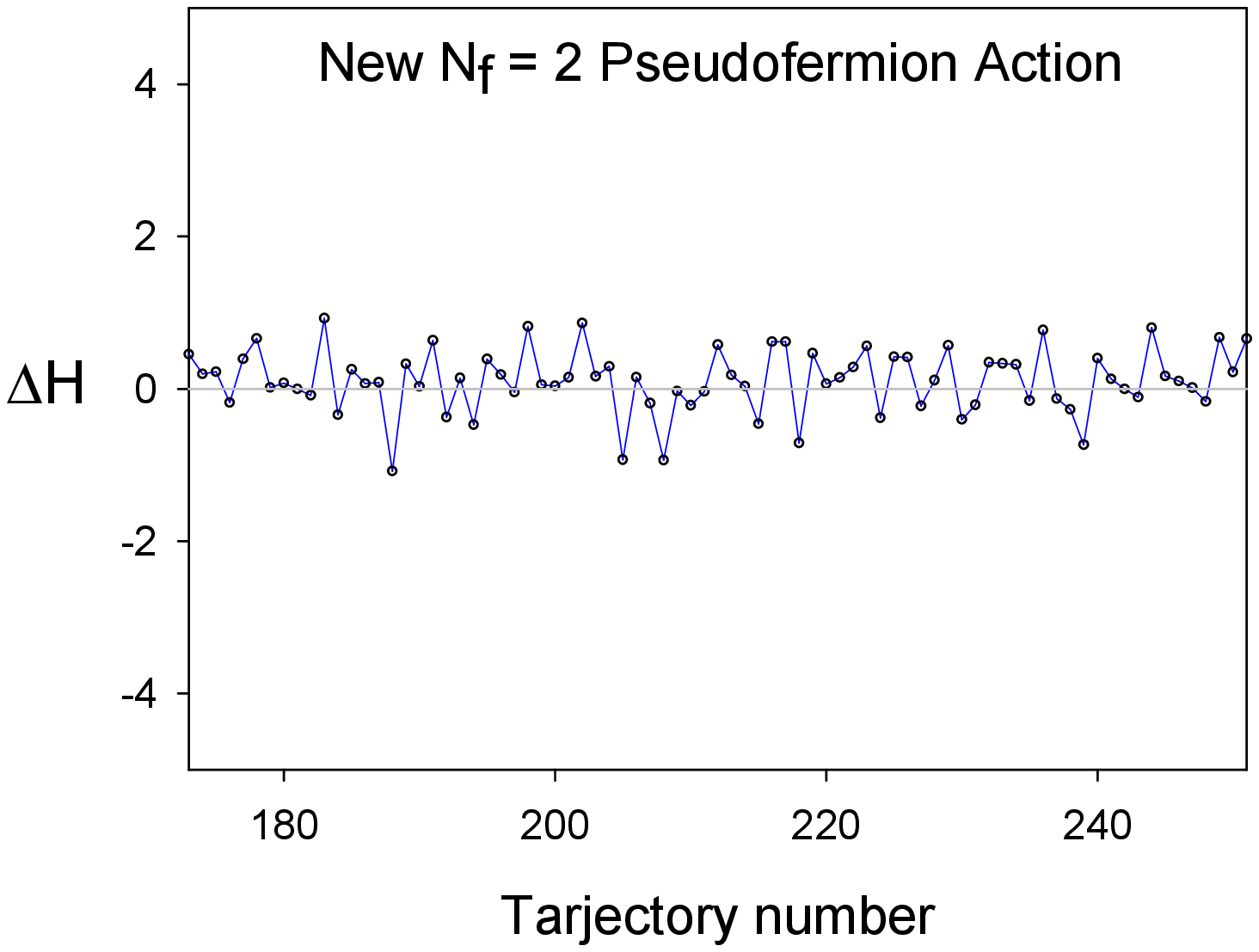}
\\ (a) & (b)
\end{tabular}
\caption{The change of Hamiltonian $ \Delta H $ versus the trajectory in the HMC simulations 
of the lattice gauge thoery with $ N_f=10 $ massless M\"obius DWF, for 
(a) the traditional action, and (b) the new action.
The line connecting the data points is only for guiding the eyes.}
\label{fig:HMC_dH}
\end{center}
\end{figure*}

In Fig. \ref{fig:HMC_dH}, we plot the change of Hamiltonian $ \Delta H $ versus the 
HMC trajectory after thermalization.
For (a) with the traditional action,  
the fluctuation of $ \Delta H $ is small, without large spikes in all trajectories. 
On the other hand, for (b) with the new action,  
the fluctuation of $ \Delta H $ is slightly larger than that in (a) with the traditional action.
Consequently, the acceptance rate of (b) with the new action is $ 0.823(43) $, 
slightly lower than that $ 0.846(41) $ of (a) with the traditional action, as 
summaried in Table \ref{tab:HMC_summary}. 
By measuring the expectation value of $ \Delta H $, we can obtain  
the theoretical estimate of the acceptance rate for HMC, 
$ P_{\rm acc} = {\rm erfc} \left( \sqrt{ \left< \Delta {\cal H} \right>}/2 \right)$ \cite{Gupta:1990ka}, 
which can be compared with the measured acceptance rate. 
As shown in Table \ref{tab:HMC_summary}, 
the measured acceptance rate is consistent with the theoretical acceptance rate 
for both cases with the traditional and the new actions. 

Next, we measure the expectation value of $ \exp(-\Delta H) $, to check whether 
it is consistent with the theoretical formula $ \left< \exp(-\Delta H) \right> = 1 $
which follows from the area-preserving property of the HMC simulation \cite{Creutz:1988wv}.
For both cases with the traditional and the new actions, 
the measured values are $ 1.011(35) $ and $ 0.983(54) $ respectively, 
in good agreement with the theoretical expectation value.

Finally, it is interesting to find out what is the renormalized coupling of this $SU(3)$ gauge theory 
with $N_f=10$ massless M\"obius DWF at $6/g_0^2 = 5.70 $. 
Performing the Wilson flow \cite{Narayanan:2006rf,Luscher:2010iy} 
with the small number ($\sim 80 $) of thermalized gauge configurations 
generated in this test, we obtain an estimate of the renormalized coupling 
$ g_c^2(L,a) \sim 4.5 $ in the finite-volume gradient flow scheme \cite{Fodor:2012td}
with $ c = \sqrt{8t}/L = 0.3 $, which is much less than the largest coupling $ g_c^2 \sim 7.0 $ 
studied in Ref. \cite{Chiu:2018edw}.

\section{Concluding Remark}

To summarize, we have constructed a new $ N_f=2 $ pseudofermion action for  
lattice gauge theory with DWF. Moreover, we demonstrate that for the $SU(3)$ gauge theory
with $N_f = 10 $ massless M\"obius DWF, the efficiency (speed $\times$ acceptance rate) 
of the HMC simulation with the new action is about 1.2 times of 
its counterpart with the traditional action.  
We expect that the gain of using the new $ N_f=2 $ pseudofermion action 
would become higher for more challenging simulations, i.e., lattice gauge theories   
with larger $N_f=2n$ (e.g., $N_f=12$) massless DWF, on larger lattices (e.g., $ 64^4 $), 
and in the stronger coupling regime. Note that for the $SU(3)$ lattice gauge theory 
with $N_f=10$ massless optimal DWF, one of us (TWC) encountered great difficulties 
(low acceptance rate and/or long simulation time) in the HMC simulation with the 
traditional $ N_f=2 $ pseudofermion action, 
on the $32^4$ lattice at $\beta=6/g_0^2=6.45$ \cite{Chiu:2017kza}.
The difficulties were circumvented by switching to the new $N_f=2$ pseudofermion action,  
and obtaining $ g_c^2(L,a) \sim 8.6 $ in the finite-volume gradient flow scheme 
with $ c = \sqrt{8t}/L = 0.3 $. Nevertheless, a detailed study 
to compare the HMC efficiencies and characteristics of this theory with  
the new and the traditional actions is beyond the scope of this paper,  
since a single GPU (e.g., Nvidia GTX-TITAN) would take more than one year 
to generate 200-300 trajectories in the HMC with the traditional $N_f=2$ pseudofermion action. 

In general, there is no guarantee that the new action would outperform 
the traditional action for any lattice field theory. 
In practice, one needs to perform numerical experiments to find out which action 
has higher HMC efficiency for the theory in question, which also depends 
on the computational platform and the algorithm implementation.
Most importantly, now we have a new pseudofermion action 
to tackle the challenging simulations of lattice gauge thoeries 
with large $N_f=2n$ massless domain-wall fermions.    


\begin{acknowledgments}

This work is supported by the Ministry of Science and Technology
(Grant Nos.~108-2119-M-003-005, 107-2119-M-003-008, 105-2112-M-002-016, 102-2112-M-002-019-MY3),
and the National Center for Theoretical Sciences (Physics Division).
We gratefully acknowledge the computer resources provided by
Academia Sinica Grid Computing Center (ASGC), and National Center for High Performance Computing (NCHC).  

\end{acknowledgments}

\end{document}